\documentclass[11pt]{aa}
\usepackage[english]{babel}
\usepackage{graphicx}
\usepackage{longtable}

\onecolumn
\def\degr{\hbox{$^\circ$}}
\def\farcs{\hbox{$^{\prime\prime}$}}
\def\farcm{\hbox{$^{\prime}$}}

\begin{document}

\title{Detailed optical spectroscopy of the B[e] star MWC\,17}

\author{V.G.~Klochkova\thanks{E-mail: valenta@sao.ru}, E.L.~Chentsov}
\institute{Special Astrophysical Observatory RAS, Nizhnij Arkhyz,  369167 Russia}
\date{\today}

\titlerunning{\it Detailed optical spectroscopy of the B[e] star MWC\,17}

\authorrunning{\it Klochkova \& Chentsov}

\abstract{Based on the data of multiple high-resolution R\,=\,60000 observations obtained at the
6-m telescope (BTA) in combination with the Nasmyth Echelle Spectrograph (NES), we studied the features 
of the optical spectrum of the star MWC\,17 with the B[e]-phenomenon. In the wavelength interval of 
4050--6750\,\AA{}, we  identiﬁed numerous permitted  and forbidden emissions, interstellar Na\,I lines, and diffuse 
interstellar bands (DIBs). Radial velocities were estimated from lines of different origin. As the systemic velocity, 
Vsys, the velocity of the forbidden emissions can be accepted: $-47$\,km/s (relative to the local standard 
Vlsr\,=$-42$\,km/s). Comparison of the obtained data with the earlier measurements allows us to conclude on 
the absence of considerable variability of spectral details.
}

\maketitle

\section{Introduction}

The hot B-type star MWC\,17\,=\,V832\,Cas is referred to the objects with the B[e] phenomenon [1, 2]. 
The B[e] phenomenon involves having a number of peculiar details in the stellar spectrum: primarily,
strong emission lines of neutral hydrogen H\,I and helium He\,I, emissions of the permitted lines of metal
ions, and low-excited forbidden lines. The second significant feature of stars with the B[e] phenomenon
is large IR flux excess due to circumstellar hot dust. However, the stars which meet these two main
criteria form a group of highly heterogeneous objects. Lamers et al. [3], having developed the classification
criteria for the stars with the B[e]-phenomenon, divided them in five subtypes. They did not refer the
star MWC\,17 to any of those subtypes, it appears in publications as a member of the most populated
B[e]-unclassified subgroup. 

A few dozens of the unclassified stars with the B[e]-phenomenon were investigated  by Miroshnichenko and coauthors. 
The review of the results obtained, including the spectral data from  the 6-m telescope, is given in [4]. 
One of the results of the spectroscopic investigation lies in the fact that  unclassified B[e]--stars are 
interacting binary systems in the broad range of luminosities. Objects with  luminosities L$\le 10^5$L$_{\odot}$ 
are separated into a new group called the objects of the FS\,CMa type. About 30\% of FS\,CMa type objects exhibit 
various signs of binarity. In [5], it is also  suggested to consider the unclassified B[e]--stars as interacting 
binaries.

The main problem in the investigation of stars with the B[e] phenomenon is the estimation of star’s luminosity. 
The luminosity would allow one to detect the evolution stage of the star and range it to one or another type of 
hot stars with similar features in the spectra. Still, in optical spectra of the stars with the B[e]-phenomenon, 
as a rule, one cannot find the absorptions which form in the stellar atmosphere conditions and could serve as 
criteria for luminosity estimation. The fact which also causes a problem is that two above mentioned distinctive 
features (spectral peculiarity and IR--flux excess) are intrinsic to several types of hot stars observed at the 
essentially different evolution stages: young Herbig Ae/Be stars, evolved massive stars of high luminosity (LBV, 
supergiants),  and far evolved low-mass stars which are close to the planetary nebula phase.

The situation with the B-type star VES\,695, which is associated with the IR-source IRAS\,00470+6429 and has 
an emission spectrum, properly illustrates the difficulties in determination of evolutionary status of 
stars with emission spectra.
For a long time, this object had been considered as a possible protoplanetary nebulae (PPN) candidate at
the stage close to the PN phase (see [6] and references therein). However, the complex study [7], carried out
with a large volume of observation material, including the echelle spectra from the 6-m telescope, resulted
in an alternate conclusion on the evolutionary status of VES\,695. Complex of the obtained fundamental
parameters (luminosity, characteristic features of energy distribution in the spectrum, binarity indications
for the star) allowed Miroshnichenko et al. [7] to refer VES\,695 to the stars of the FS\,CMa type.

MWC\,17 which is associated with the IR-source IRAS\,01441+6026 is one of the hottest stars with the
B[e]-phenomenon. It is located near the galactic plane and has the coordinates: $\alpha$(2000)\,=\,01$^h$\,47$^m$\,38.5$^s$,
$\delta$(2000)\,=+60${\degr}$\,41${\farcm}$\,57${\farcs}$,  l\,=\,129.8$\degr$, and b\,=\,$-1.4{\degr}$. 
In the optical range, MWC\,17 is quite a faint star with a magnitude V\,=\,11$\lefteqn{.}^m7$ [8]. Due to high reddening, 
the star is weaker in the B band, B\,=\,13$\lefteqn{.}^m5$ (according to the SIMBAD data), thus, its optical spectrum 
is understudied at present. Zickgraf [8] carried out the only spectroscopic study of MWC\,17 with high spectral
resolution. The spectral resolution of his data (R from 23000 to 45000) is comparable to ours, but this author used 
only small regions of the spectrum. The H$\alpha$, He\,I 5876, [NII]\,6583, [OI]\,6300, [SIII]\,6312, and [FeII]\,7155\,\AA{} 
profiles are presented in [8] graphically. Relative intensities and radial velocities are given for these lines as well as 
for HeI\,6678 and NaI\,5890, 5896\,\AA{}.

To search for possible variability of the MWC\,17 optical spectrum and to add data on its peculiarities,
we made detailed identification and necessary measurements of the parameters of spectral details
of different origin from the spectra obtained with a resolution R\,=\,60000 in the broad wavelength range 
4050--6750\,\AA{}. The observation data used is briefly described in Section\,2. Section\,3 presents information 
on the profile of spectral details found with the high-resolution spectra, their analysis, and
discussion of the results derived. In Section\,4, we deal with the position of MWC\,17 in the Galaxy, and the
main conclusions are summed up in Section\,5.

\section{Observed  data}

We use the MWC\,17 spectra obtained on the 6-m telescope with the Nasmyth Echelle Spectrograph
(NES) [9]. The spectrograph equipped with a CCD of 2048$\times$2048 elements and an image slicer [10]
provides R\,=\,60000 in the wavelength range  of 3500--6800\,\AA{}. One-dimensional spectra were extracted 
from two--dimensional echelle images using the modified~[11] MIDAS Echelle context.
Table\,1 shows the dates of obtaining the spectra and the detected spectral intervals. Correction and control
of instrumental coordination of spectra of the star and the lamp with a hollow cathode were fulfilled with
the telluric lines [OI], O$_2$, and H$_2$O. The procedure of radial velocity Vr determination from the spectra
obtained with NES and error sources are described in more detail in [12]. The root-mean-square error of the
measurements is Vr$\le$0.8\,km/s for single narrow line.

\begin{table}[ht!]
\caption{Dates of obtaining the spectra used and the detected spectral interval} 
\medskip
\begin{tabular}{c| c}
\hline
 Date   &   $\Delta\lambda$,\,\AA{}  \\  
\hline                       
  15.11.2005  &   5275$\div$6735     \\
  15.01.2006  &   4570$\div$5980     \\
  16.01.2006  &   4570$\div$5980     \\
  14.03.2006  &   4050$\div$5450     \\
  02.09.2006  &   5275$\div$6735     \\
\hline 
\end{tabular}
\label{spectra}
\end{table}

\section{Main types of line profiles}

As early as in the first publication, when were mentioned some features of the MWC\,17 spectrum [13], a remarkably 
high intensity of emissions was noted. Intensity variations, from peaks of the strongest HI, [OI] emissions to the NaI 
absorption cores, and therefore lowering of the signal-to-noise ratio, attain three orders of magnitude in the MWC\,17
spectrum! Weak emissions, which are hardly distinguishable from noise, make it difficult to generate the continuum. 
Absorptions, by contrast, (except for depressions in the emission lines) are presented by the interstellar NaI\,(1) 
lines, the strongest of DIBs and, possibly, by the H$\delta$ photospheric wings.
With regard to these circumstances, we detected only small variations of the intensities of the lines and
profile shapes and positions with time. The strongest emissions were on September~2,~2006; the residual
intensities of the emission peaks in this spectrum are on the average 20\% higher than in January and 30\%
higher than in March of the same year.

\begin{figure}[hbtp]
\includegraphics[angle=0,width=0.6\textwidth,bb=20 40 710 520,clip=]{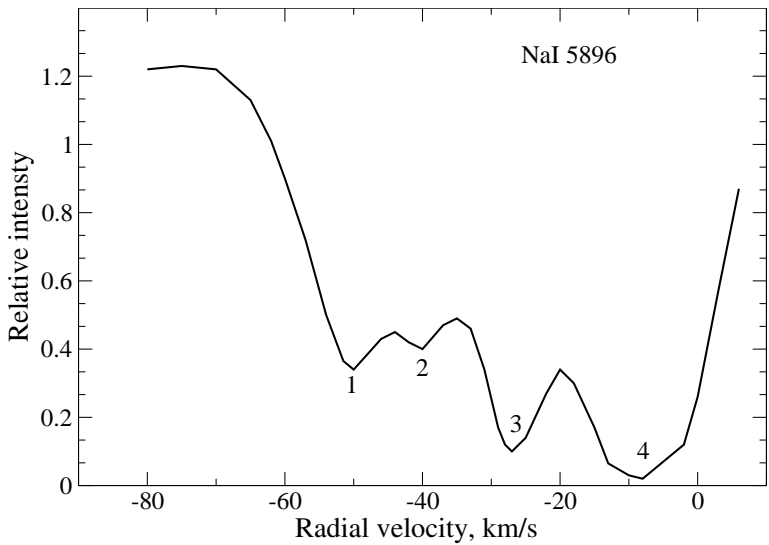}
\caption{Profile of the NaI\,5896\,\AA{} line in the spectrum of MWC\,17 averaged over all available spectra. 
      Numbers 1--4 denote the positions of four profile components from Table\,3.}
\label{fig1}                                        
\end{figure}

Heliocentric radial velocities, Vr, were found for the profiles on the whole or for separate details by
coincidence of their direct and mirror images. When comparing Vr for the same lines in the spectra obtained 
on different dates, it should be taken into account that apart from random errors, systematic errors of the 
order of 1\,km/s are also possible. Table\,2 shows the examples of the comparison of our data with each other 
and with the data from [8]. The profile shape which is complex and changing from line to
line does not allow us to confine ourselves to intensity and a single value of radial velocity only. These and
additional profile parameters are collected in Tables\,2, 3, 4, and 5. In these tables, {\it r} is the residual intensity
of an emission peak or an absorption core, V/R is the ratio of residual intensities of the blue and red components 
for double--peaked emissions. The velocities are rounded to whole km/s, they correspond to: 

Vr -- the upper part of the whole profile for the emission or the lower part of the absorption component;

V$_{em}$ -- the emission peaks of a double-peaked profile;

V$_{abs}$ -- the depression in the upper part of the emission line profile;

V(r/2) -- the shortwave and longwave (hereinafter ``blue'' and ``red'') slopes of the emission line profile
       at the half maximum intensity;

V(r$\approx$1) -- the blue and red boundaries of the emission at the bottom of the profile.

The results from [8] are given in italics. Table\,3 gives the velocities measured from the NaI\,(1) absorption components. 
       
\begin{table}[hbtp]
\caption{Profile parameters (see in the text) for some lines and dates according to our measurements and Zickgraf\`\,s
         data~[8] (in italics). Heliocentric velocities Vr are in km/s. Uncertain measurements are marked with a colon}
\medskip                                                                           
\begin{tabular}{r @{{\vrule width 1pt}}  c  @{{\vrule width 1pt}}  c @{{\vrule width 1pt}} 
                c @{{\vrule width 1pt}}  c| r @{{\vrule width 1pt}} l @{{\vrule width 1pt}} c| c @{{\vrule width 1pt}} l| c}
\hline                                                                                                                                          
Date &\hspace{0.2cm} r \hspace{0.2cm} &\hspace{0.2cm}  V/R \hspace{0.2cm}  & \hspace{0.2cm} Vr\hspace{0.2cm} & 
        \multicolumn{2}{c @{{\vrule width 1pt}}}{\footnotesize V$_{em}$}
        &  V$_{abs}$   &\multicolumn{2}{c@{{\vrule width 1pt}} }{\footnotesize V(r/2)} & \multicolumn{2}{c}{\footnotesize V(r$\approx$1)} \\
\hline                                 
\multicolumn{11}{c}{${\rm [OI]}$\,6300}    \\   
15.11.05     &      33  &     0.94  & $ -48$  &  $-59$  &  $-38$ &    $-51$ &     $-74$   &   $-23$   & $-92$  &      0    \\    
 2.09.06     &      33  &     0.97  & $ -48$  &  $-57$  &  $-40$ &    $-49$ &     $-75$   &   $-24$   & $-95$  &     $-5$     \\    
             &{\it  25} &{\it 0.91} &$-${\it47}:&$-${\it59}&$-${\it39}&$-${\it51}&$-${\it72}:&$-${\it21}:&$-${\it92}:&$-${\it2}: \\ 
\multicolumn{11}{c}{$\rm [FeII]$\,5334}  \\       
15.11.05     &      3.7 &           & $ -48$  &         &        &          &     $-72$   &   $-26$   & $-95$: &     $ 3$:   \\    
15.01.06     &      3.7 &           & $ -46$  &         &        &          &     $-71$   &   $-24$   & $-94$: &     $-1$:   \\    
16.01.06     &      4.2 &           & $ -47$  &         &        &          &     $-72$   &   $-23$:  & $-95$: &     $ 0$:   \\    
14.03.06     &      3.7 &           & $ -46$  &         &        &          &     $-72$   &   $-26$:  & $-93$: &     $-5$:   \\    
02.09.06      &     4.9 &           & $ -47$  &         &        &          &     $-72$   &   $-28$   & $-94$: &     $-5$:   \\    
\multicolumn{11}{c}{${\rm [NII]}$\,5755}   \\     
15.11.05     &      16  &     0.84  &  $-47$  &  $-65$  &  $-32$ &    $-52$ &     $-85$:  &   $-16$   & $-125$:&       5  \\     
15.01.06     &      14  &     0.82  &  $-47$  &  $-64$  &  $-31$ &    $-53$ &     $-86$:  &   $-15$   & $-125$:&       6  \\     
16.01.06     &      15  &     0.80  &  $-47$  &  $-64$  &  $-32$ &    $-55$ &     $-85$:  &   $-15$:  & $-124$:&      10: \\     
02.09.06     &      21  &     0.80  &  $-48$  &  $-64$: &  $-32$ &    $-52$ &     $-83$:  &   $-17$   & $-120$:&       7:  \\    
\multicolumn{11}{c}{FeII\,5169} \\    
15.01.06     &     5.7  &     0.74  &  $-42$  &  $-55$: &  $-36$ &   $-50$: &     $-69$:  &   $-16$   & $ -98$:&       2: \\    
16.01.06     &     6.7  &     0.72  &  $-41$  &  $-55$: &  $-37$ &   $-51$: &     $-71$:  &   $-18$   & $-102$:&       0: \\    
14.03.06     &     4.6  &     0.78  &  $-43$  &  $-57$: &  $-35$ &   $-51$: &     $-71$:  &   $-19$:  & $ -97$:&       5: \\    
\multicolumn{11}{c}{FeII\,5316}\\
15.11.05     &     13   &     1.00  &  $-51$  & $-69$   &  $-35$ &   $-50$  &     $-88$   &   $-13$   & $-112$ &    11     \\    
15.01.06     &     11   &     0.95  &  $-51$  & $-68$   &  $-37$ &   $-51$  &     $-88$   &   $-14$   & $-115$:&    12:    \\    
16.01.06     &     12   &     0.94  &  $-51$  & $-69$   &  $-36$ &   $-52$  &     $-88$   &   $-14$:  & $-111$:&    14:    \\    
14.03.06     &     10   &     1.07  &  $-52$  & $-70$   &  $-33$ &   $-52$: &     $-90$:  &   $-13$:  & $-115$:&    14:    \\    
02.09.06     &     15   &     1.12  &  $-54$  & $-70$   &  $-34$ &   $-49$  &     $-90$   &   $-15$   & $-108$:&    11     \\    
\multicolumn{11}{c}{FeII\,6318} \\
15.11.05     &     8.5  &     0.94  &  $-50$  &         &  $-34$ &   $-50$  &     $-86$   &   $-13$   & $-105$:&      8:  \\   
02.09.06     &     9.1  &           &  $-53$  &         &  $-39$ &          &     $-89$   &   $-15$   & $-105$:&      8:  \\   
\multicolumn{11}{c}{${\rm [SIII]}$\,6312}\\ 
15.11.05     &     4.0  &     0.73  &  $-47$: &         &  $-25$ &          &     $-81$:  &   $-12$   & $-118$:&     10:  \\   
02.09.06     &    5.3  &     0.68   &  $-46$: &  $-66$: &  $-28$ &  $-50$   &     $-80$:  &   $-14$   & $-122$:&       6: \\   
             & {\it 3.1}&{\it 0.76} &$-${\it50}:&$-${\it70}&$-${\it25}&$-${\it44}&$-${\it88}:&$-${\it12}:&$-${\it123}:& {\it4:}  \\ 
\multicolumn{11}{c}{HeI 5876} \\
15.11.05     &      7   &     0.82  & $-54$   &  $-78:$ & $-28$  &  $-52$:  &    $-108$   &    $-6$   & $-145$:&     25   \\    
15.01.06     &      6   &     0.76  & $-50$   &  $-70$: & $-27$  &  $-57$:  &    $-110$   &    $-3$   & $-142$:&     33:  \\    
16.01.06     &      6   &     0.75  & $-48$:  &         & $-26$  &          &    $-106$:  &    $-3$:  & $-150$:&     33   \\    
02.09.06     &      9   &     0.75  & $-56$   &  $-80$: & $-33$  &          &    $-110$   &    $-11$  & $-140$:&     20:  \\    
             & {\it 6}  & {\it 0.68}&$-${\it50}:&$-${\it78}&$-${\it24}&$-${\it57}&$-${\it105}:&{\it0}:&$-${\it135}:&{\it30}:  \\ 
\multicolumn{11}{c}{HeI\,6678 } \\
15.11.05     &      2.3 &     0.84  & $-51$   &  $-72$  &  $-23$ &  $-60$:  &    $-106$:  &    $-1:$  & $-138$:&    20:  \\      
02.09.06     &      2.7 &     0.78  & $-54$:  &         &  $-28$:&          &    $-105$:  &    $-2:$  & $-135$:&    22:  \\      
             &          &           &         &$-${\it67}&$-${\it20}&$-${\it53}&          &           &        &           \\    
\multicolumn{11}{c}{H$\alpha$}\\
15.11.05     &          &           &         & $-105$: &  $-10$ &    $-54$ &             &           &        &           \\    
02.09.06     &          &           &         & $-100$: &  $-13$ &    $-55$ &             &           &        &           \\    
             &          &           &         &$-${\it101}&$-${\it11}&$-${\it59}&         &           &        &           \\    
\hline                                                                                                                                 
\end{tabular}
\label{Profiles}
\end{table}

As follows from Table\,2, the variations of radial velocities for strong emissions [OI] with distinct profiles in our 
two spectra, and the discrepancy with the data from [8] do not go beyond the measurement error.
The same can be said about the weaker [NII] and [FeII] emissions. Velocity variations from spectrum to
spectrum for other lines are notable. They are mainly due to the variations in the red regions of the profiles
(the shifts of their red slopes and peaks). The shifts for the [SIII] and FeII emissions are about 2--3\,km/s,
for HeI\,5876\,\AA{} they attain 5\,km/s.

Red regions are also most variable in the complex absorption/emission profiles of the NaI doublet. Figure\,1 shows the 
result of averaging the profiles of the NaI\,5896\,\AA{} line using several our spectra. The emission component is presented 
in this figure only by its blue region (Vr of $-60$\,km/s), its red region is overlapped by interstellar extinction. 
Table\,3 shows  that three blue components of the NaI\,5896\,\AA{} line keep their positions while the red component 
is shifted on November~15,~2005 by $-$(5--6)\,km/s relative to its position in 2006.

The double-peaked H$\alpha$ emission profile is of type III according to the Beals classification~[14].
The position of emission peaks Vr$\approx-100$ and $-12$\,km/s as well as the position of the absorption
Vr$\approx -55$\,km/s are almost constant both in our two spectra of 2005, 2006 and in the earlier spectrum
by Zickgraf~[8]. One can see the change in the strongest emission intensities as compared with the data from [8] 
(see Table A1 in that paper). Both the [OI]\,6300\,\AA{} and [SIII]\,6312\,\AA{} lines with an excitation potential 
of 3.3\,eV are 30\% stronger in our spectrum than those in~[8]. The HeI\,5876\,\AA{} line has scarcely varied 
as compared with the measurements by Zickgraf~[8].

\begin{table}[ht!]
\caption{Radial velocities Vr measured for the components of the NaI doublet D-lines for certain dates. 
         Zickgraf’s measurements are given in italics~[8] } 
\medskip
\begin{tabular}{ c| c|  c| c| c}
\hline
   Дата       & \multicolumn{4}{c}{Vr, km/s}  \\ 
  \cline{2-5}   
              &    1   & 2   & 3   & 4   \\
\hline   
\underline{NaI\,5890}&    &          &          &       \\  
  15.11.2005  &$-52$      & $-40$    & $-25$    & $-14$  \\ 
  15.01.2006  &$-50$      & $-38$    & $-27$    & $ -8$ \\  
  16.01.2006  &$-50$      & $-38$    & $-27$    & $ -9$ \\  
  02.09.2006  &$-50$      & $-38$    & $-28$    & $ -9$ \\  
              &$-${\it52} &$-${\it41}&$-${\it25}&$-${\it13} \\   
\underline{NaI\,5896}&    &          &          &       \\       
15.11.05      &$-50 $     &$-40$     & $-25$    & $-12$ \\     
15.01.06      &$-50 $     &$-40$     & $-27$    & $ -9$ \\     
16.01.06      &$-50 $     &$-41$     & $-27$    & $ -8$ \\     
02.09.06      &$-53 $     &$-42$     & $-26$    & $ -9$ \\     
              &$-${\it49} &$-${\it42}&$-${\it25}&$-${\it11} \\     
\hline 
\end{tabular}
\label{Na}                                      
\end{table}

\begin{figure}
\vbox{
\includegraphics[angle=0,width=0.45\textwidth,bb=30 30 570 810,clip=]{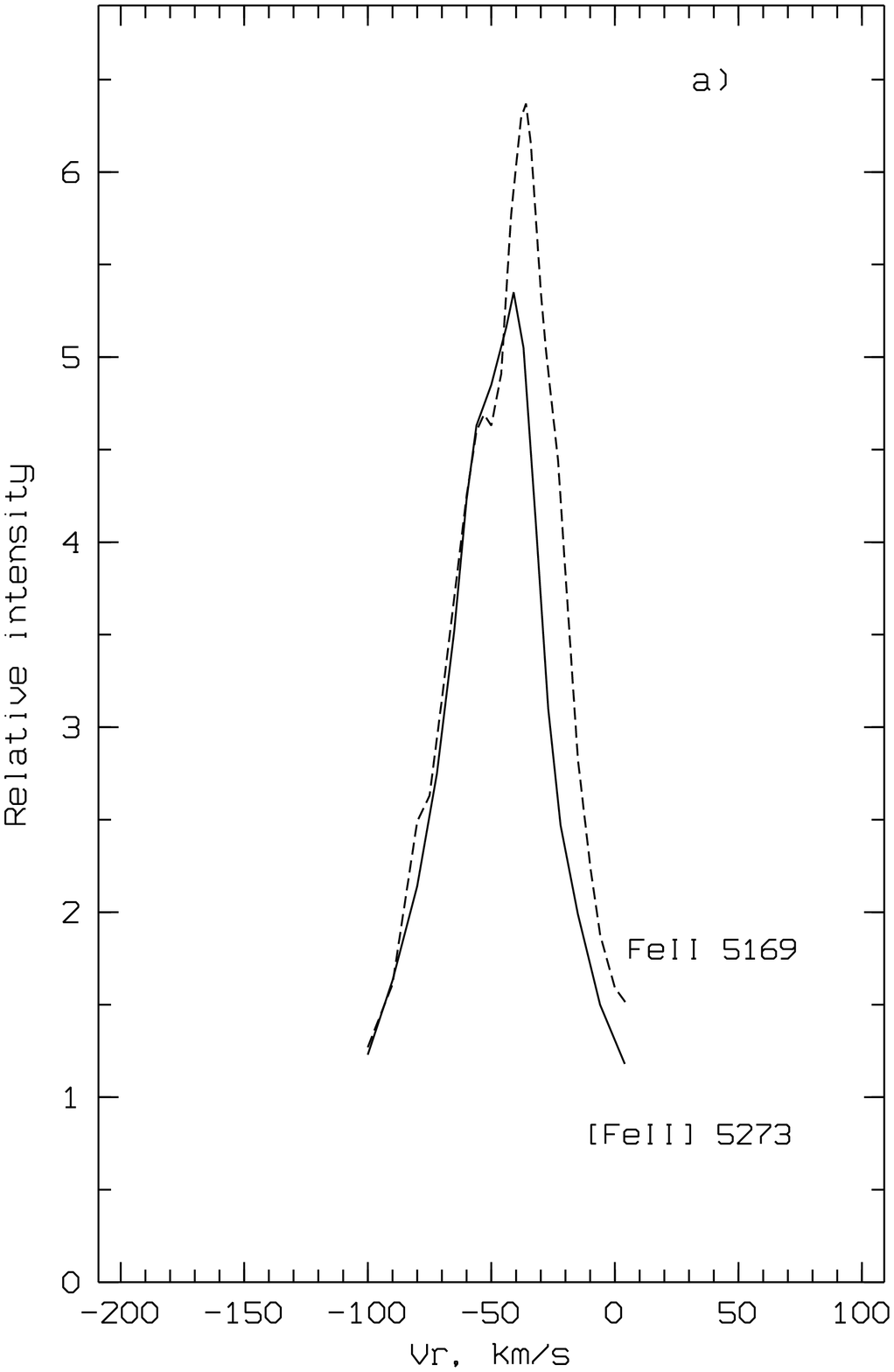}
\includegraphics[angle=0,width=0.45\textwidth,bb=30 30 570 810,clip=]{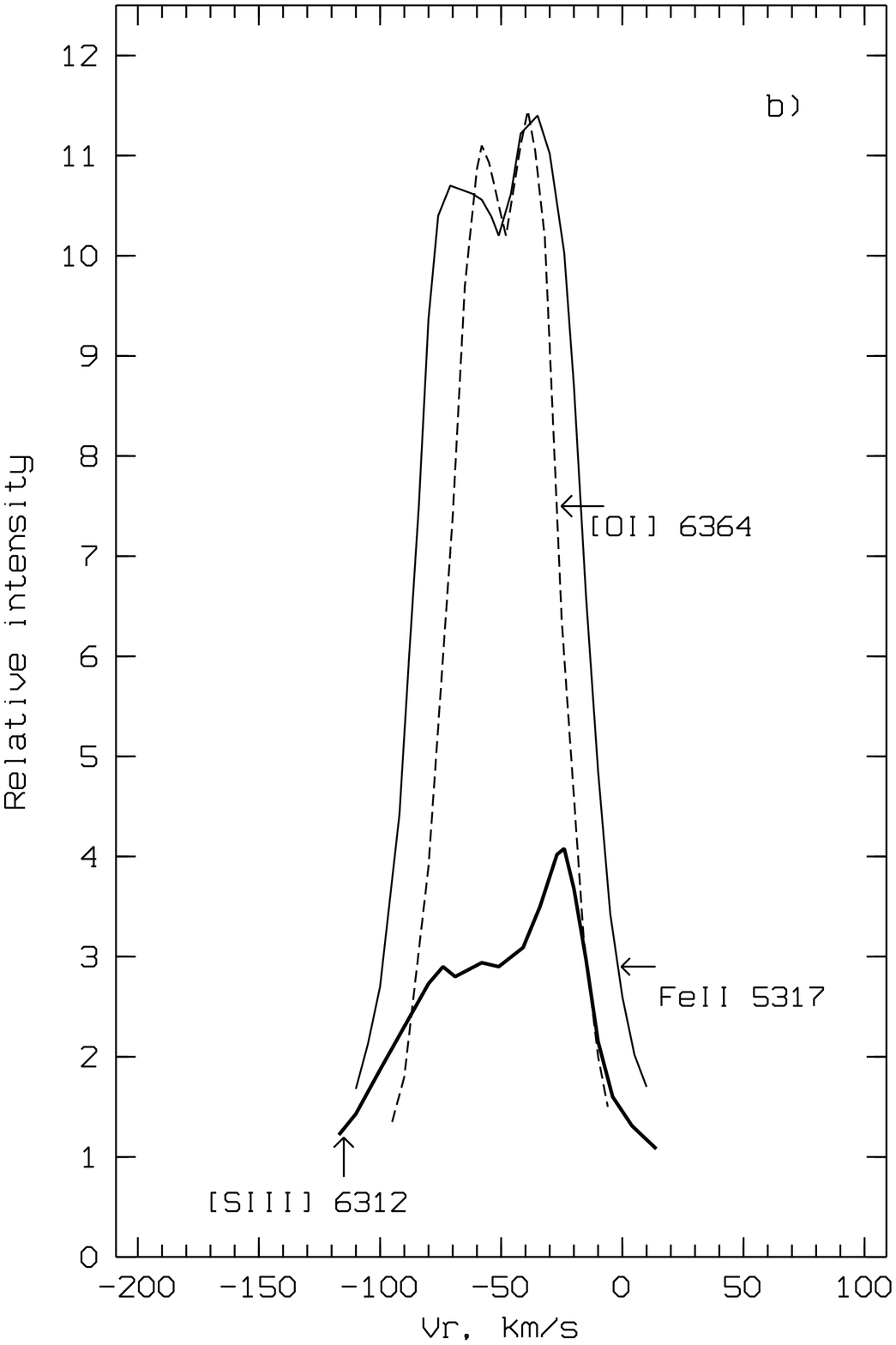}
\includegraphics[angle=0,width=0.45\textwidth,bb=30 30 570 810,clip=]{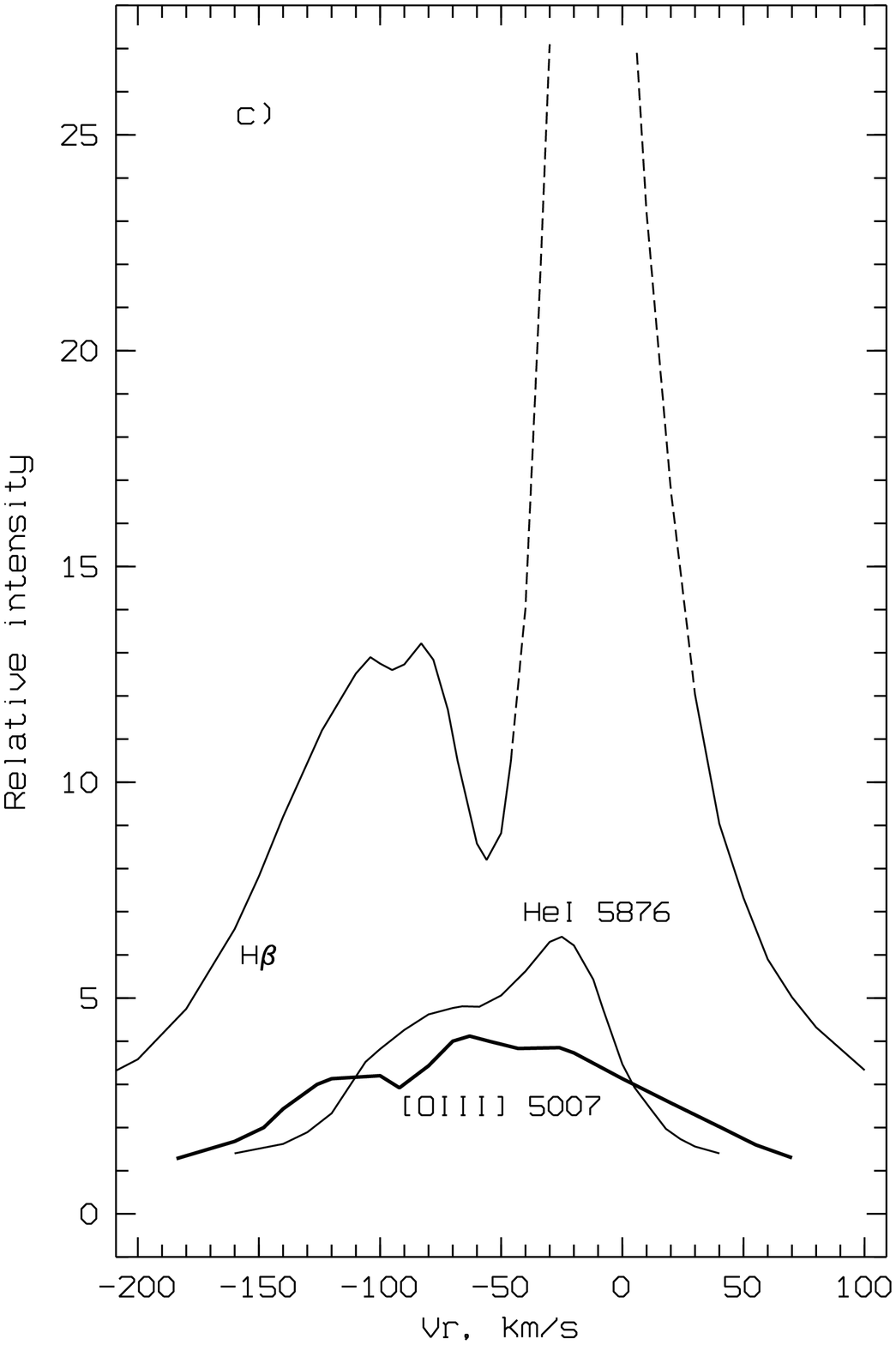}
}
\caption{Profiles of the representative lines in the MWC\,17 spectrum: (a) FeII\,5169\,\AA{} (dashed line) and [FeII]\,5273\,\AA{} (solid);
  (b) FeII\,5317\,\AA{} (thin solid line), [SIII]\,6312\,\AA{} (thick solid), and [OI]\,6364\,\AA{} (dashed); (c) [OIII]\,5007\,\AA{} 
  (thick solid line), HeI\,5876\,\AA{} (thin solid), and H$\beta$ (thin solid, prolonged with the dashed line at high intensities).} 
\label{fig2}                                        
\end{figure}

Table\,4 and Fig.\,2 illustrate the hierarchy of emission profiles by width and degree of asymmetry: both
increase from top lines of the table to the bottom. The velocities in this table are averaged for the groups of
lines with a similar profile shape. They can be useful not only in estimating the structure and kinematics
of the MWC\,17 envelope but also in controlling the line identification accuracy. As mentioned in previous
studies of MWC\,17 (and which is commonly characteristic of B[e]--stars), the narrowest lines in the optical
spectrum are the forbidden emissions [FeII] and [OI]. Both are almost symmetrical. The former ones have
sharper peaks (bifurcation is notable only in some lines and in some our spectra), the latter ones have
more distinctly forked peaks. The red peak of the [OI] profile is stronger than the blue one.

In accordance with the results from [8], such a ratio of peak intensities is amplified for the HI, HeI, and
[SIII] lines. According to our data, it is also typical of [NII], [OIII], [FeIII] and generally of most emissions.
An exception is provided by the FeII lines of low  excitation  (e.g., FeII\,(49, 48)\,5316\,\AA{}, see Table\,2), 
the blue component of which can be stronger than the red one.

The permitted FeII emissions are notably broader than the forbidden ones, and they are slightly asymmetric: the blue 
slope is steeper than the red one. The low and high-excitation FeII lines (about 3 and 10\,eV respectively) noticeably 
differ by the shape of the profile upper part: it is close to rectangular for the latter ones (e.g., the FeII\,6318\,\AA{} 
line proﬁles in Table\,2), the central depression is not distinct. Difference in widths of the permitted and forbidden 
emissions indicate that these groups of lines origin in different physical conditions.

The members of the 42-nd FeII multiplet, which is the strongest in the visible spectrum, stand out for
their profile shape. They are as broad as the other FeII emissions but abruptly get narrow toward the tops
due to intensity decrease in the blue half of the line. As can be seen from Table\,4 and Fig.\,2, the positions
of the red emission peak and the whole red slope of the profile are almost the same as in other Fe\,II lines; 
the blue peak is weak and shifted toward the red region, it is faintly visible on the blue slope which is much
smoother than the red one. Such asymmetry (the blue peak is not distinct and is weaker than the red one,
the blue slope is smoother than the red one) is as well characteristic of emissions of forbidden ions [NII],
[SIII], and [FeIII] which are broader at the bottom. It also remains in the broadest (exclusive of those of
hydrogen) HeI and [OIII] lines. Among the model profiles~[8], those are the most similar to the observed
spectrum of MWC 17 which are computed for a tilted disk (i$\approx 45{\degr}$), while Zickgraf~[8] considers 
that i$\approx 80{\degr}$ for MWC\,17.

\begin{table}[ht!]
\caption{Mean heliocentric radial velocities for emission groups in the MWC\,17 spectrum with a similar profile shape.
          Designations are the same as for Table\,2. Uncertain measurements are marked with a colon } 
\medskip
\begin{tabular}{c| c| c  c| c| c  c| c r}
\hline
  Line type                &   Vr  &\multicolumn{2}{c|}{V$_{em}$} &  V$_{abs}$  & \multicolumn{2}{c|}{V(r/2)} & \multicolumn{2}{c}{V(r$\approx$1)} \\
\hline
${\rm [FeII]}$             &  $-47 $  &   $ -53$: &  $-42$: &   $-50$   & $ -72$ & $ -26$  & $ -95$  & $ -5$: \\
${\rm [OI]}$, ${\rm [SII]}$&  $-48 $  &   $ -58$  &  $-39$  &   $-50$   & $ -74$ & $ -24$  & $ -93$  & $ -4$  \\
FeII  low exc.             &  $-51 $  &   $ -71$  &  $-34$  &   $-49$   & $ -89$ & $ -11$  & $-105$: & $  9$  \\
FeII  high exc.            &  $-51 $  &   $ -63$: &  $-36$: &   $-49$:  & $ -87$ & $ -14$  & $-105$: & $  9$: \\
FeII(42)                   &  $-42 $  &   $ -55$: &  $-36$  &   $-50$:  & $ -70$ & $ -18$  & $-103$: & $  5$: \\
${\rm[NII]}$, ${\rm[SIII]}$&  $-47 $  &   $ -64$  &  $-30$  &   $-52$   & $ -83$ & $ -15$  & $-120$: & $  8$:\\
HeI                        &  $-53:$  &   $ -76$: &  $-28$: &   $-55$:  & $-108$:& $  -6$: & $-142$: & $ 25$:\\
${\rm [OIII]}$             &  $-55:$  &   $-100$: &  $-50$: &   $-78$:  & $-130$:& $  20$: & $-180$: & $ 90$:\\
H$\alpha$, H$\beta$        &  $-52:$  &   $ -95$: &  $-12$  &   $-55$   &        &         &         &        \\
\hline                                                                                                                                 
\end{tabular}
\label{Vr}
\end{table}

Table\,5 shows the identified lines and their parameters. As the discrepancies in the parameters
from spectrum to spectrum for most lines are small, each line is presented here by a single set of the parameters. 
In the wavelength interval 4050--4570\,\AA{}, the parameters were obtained from the spectrum of March~14, 2006; 
in longer wavelength intervals, they were found by averaging two or more spectra. As  follows from Table\,5, 
we have not detected any absorptions in the MWC\,17 spectrum with all the high-quality material in the broad 
wavelength interval. Using spectra with a resolution limit of 1.3\,\AA{}, Jaschek and Andrillat~[15] found 
several absorptions forming in the stellar atmosphere: AlII\,6066, SII\,6102, NII\,6632, and CII\,6800\,\AA{}.
We have not detected the ﬁrst three absorptions, which could occur in our spectral interval.

\section{Location of MWC\,17 in the Galaxy}

If not particularly the radial velocity of the star itself (it may be peculiar), then at least the presence of the 
blue-shifted components in the interstellar NaI\,(1) lines with velocities of up to $-50$\,km/s  (there is even a weak 
component with a velocity of about $-60$\,km/s) says that MWC\,17 is located in the Perseus Arm or behind it. 
The distance to MWC\,17 can be estimated by different methods, but the peculiarity of the object make all them unreliable.
\begin{itemize}
\item Spectrophotometric parallax. The absence of photospheric absorptions in the star’s spectrum
makes the accepted (B-V)$_0$ color index uncertain. Martel and Gravina~[16] give the following for MWC\,17: 
V\,= 11$\lefteqn{.}^m$66,   (B-V)\,=\,0$\lefteqn{.}^m$42;  along with this, for early subtype B we obtain
the color excess E(B-V)$\approx 0\lefteqn{.}^m$65 and visual brightness extinction Av$\approx 2\lefteqn{.}^m$5. 
Let us notice that we obtain a smaller excess E(B-V)$\approx 0\lefteqn{.}^m4$ using the measured equivalent width 
of the interstellar band W(5780)$\approx$0.2\,\AA{} and correlation W(5780)--E(B-V) from~[17]. 

Even with Mv$\approx-4^m$ (such a luminosity would rather correspond to a star from the upper part of the main
sequence than to a supergiant), d$\approx4$\,kpc. It is the lower limit of the MWC\,17 distance. However, the
estimates of both the star’s color and its spectral type are unreliable until the distorting influence of 
emissions is taken into account and photospheric lines are detected. However, if we bear in mind that according to 
Miroshnichenko~[18], the visual magnitude of the star V$\approx 13^m$, then the distance to it will be considerably greater.

\item  According to [19], the interstellar extinction toward MWC\,17 reaches 2$\lefteqn{.}^m$5 at a distance of 1\,kpc
and remains almost the same up to 3--4\,kpc. If we assume the above estimated Av$\approx 2\lefteqn{.}^m$5 for MWC\,17, 
we will obtain only a limitation on the object’s distance, d$>$1\,kpc.

\item  Among the velocities given in Table\,4, the closest to Vsys can be the velocities of the forbidden emissions: 
$-47$\,km/s (Vlsr\,=$-42$\,km/s). By Brand and Blitz~[20], a distance d$\ge$2\,kpc corresponds to this value.
\end{itemize}

There is an important additional point: in the sky, MWC\,17 is located at the southern boundary
of the compact association Cas\,OB\,8 (d$\approx$2.9\,kpc from~[21]). For the member of this association
HD\,9311, Vr\,=$-42$\,km/s; the structure of the interstellar NaI lines in its spectrum is similar to that in 
the MWC\,17 spectrum (Vr$\approx-(65\div50)$,  $-14$\,km/s~[22]). Obviously, the question on possible
membership of MWC\,17 in Cas\,OB\,8 is a key one for determination of the distance and origin of the object.
Meanwhile, we have to assume that the distance to MWC\,17 can be of the order of both 4--5\,kpc and 2--3 
kpc --- in the case there is some additional (possibly circumstellar) extinction.

The profile of the NaI\,D lines in the MWC\,17 spectrum is close to that observed in the spectrum
of the hot B-type star -- the optical component of the IR-source IRAS\,00470+6429, which is located
near the galactic plane and has the coordinates  $\alpha(2000)$\,=\,00$^h$\,50$^m$\,06$^s$, $\delta(2000)$\,=\,+64${\degr}$\,45$\farcm$\,35$\farcs$,
l\,=\,122$\lefteqn{.}{\degr}8$, and b\,=\,1$\lefteqn{.}{\degr}9$. In the optical spectrum of its central star, there are two interstellar components
of the NaI\,D lines~[7]. Heliocentric radial velocities in the IRAS\,00470+6429 spectrum (about $-62$ and $-13$\,km/s) coincide with those in 
the MWC\,17 spectrum. In the spectra of both objects, the positions of diffuse interstellar bands are also similar:
Vr(DIB)$\approx -14$\,km/s.

\section{Conclusions}

High quality of the spectral data which we used allowed us to identify numerous details in the spectrum of MWC\,17. 
The thorough search has not led to detection of any absorptions in the  spectrum 
which emerge in the stellar atmosphere. Out of absorption details, only interstellar components of the
NaI\,D lines and DIBs were identified. The DIBs are weak, the  equivalent widths are W(5780)$\approx0.2$\,\AA{} 
and W(5797)$\approx0.15$\,\AA{}.

As a systemic velocity, Vsys, the velocity for forbidden emissions can be accepted: $-47$\,km/s 
(relative to the local standard Vlsr\,=$-42$\,km/s).

Comparison between our results and the measurements of Zickgraf~[8] allows us to draw an inference
on the absence of significant variability of spectral details.

We noticed variations of radial velocities from spectrum to spectrum which are small and can hardly
give sure evidence on spectral binarity of MWC\,17. In order to solve this problem, observations of this star
with high positional accuracy should be continued. Taking into account high intensity of the emission
details in the spectrum of this faint star, such observations can be also made on telescopes of a
moderate diameter.

\section*{Acknowledgements}

The authors are grateful to V.E.\,Panchuk and M.V.\,Yushkin for considerable assistance with the
observations on the BTA. The study was fulfilled   with the financial support of the Russian Foundation for Basic
Research (project 14--02--00291\,a). The observations on the 6-m telescope of the Special Astrophysical
Observatory was conducted  with the  support of the Ministry of Education and Science of 
the Russian Federation (agreement No.\,14.619.21.0004, project ID\,RFMEFI61914X0004). The astronomical 
databases SIMBAD and ADS were used in the study.

\newpage

\section*{REFERENCES}

\begin{enumerate}

\item D.A.~Allen and J.P.~Swings, Astron. \& Astrophys. \textbf{47} 293 (1976).

\item  P.S.~The, D.~de~Winter, \& M.R.~Perez, Astron. \& Astrophys. Suppl. \textbf{104} 315 (1994).

\item  H.J.G.L.M.~Lamers, F.J.~Zickgraf, D.~de~Winter, et al., Astron. \& Astrophys. \textbf{340} 117 (1998).

\item  A.S.~Miroshnichenko, Astrophys. J. \textbf{667} 497 (2007).

\item  M.~Kraus, M.~Borges Fernandes, \& O.~Chesneau, ASP Conf. Ser. \textbf{435} 395 (2010).

\item  D.M.~Kelly \& B.J.~Hrivnak, Astrophys. J. \textbf{269} 1040 (2005).

\item  A.S.~Miroshnichenko, E.L.~Chentsov, V.G.~Klochkova, et al., Astrophys. J. \textbf{700} 209 (2009).

\item  F.--J.~Zickgraf, Astron. \& Astrophys. \textbf{408} 257 (2003).

\item  V. Panchuk, V. Klochkova, M. Yushkin, \& I. Najdenov, J. Optical Technology \textbf{76} 87 (2009).

\item  V.E.~Panchuk, M.V.~Yushkin, \& I.D.~Najdenov, Preprint No.\,179 (Special Astrophysical Observatory, Nizhny Arkhyz, 2003).

\item  M.V.~Yushkin \& V.G. Klochkova, Preprint No.\,206 (Special Astrophysical Observatory, Nizhny Arkhyz, 2005).

\item  V.G.~Klochkova, V.E.~Panchuk, M.V.~Yushkin, \& D.S.~Nasonov, Astrophysical Bulletin \textbf{63} 386 (2008).

\item  P.W.~Merrill \& C.G.~Burwell, Astrophys. J. \textbf{78} 87 (1933).

\item  C.S.~Beals, Publ. Dominion Astrophys. Obs. \textbf{9} 1 (1953).

\item  C.~Jaschek \& Y.~Andrillat, Astron. \& Astrophys. Suppl. \textbf{136} 53, (1999).

\item  M.T.~Martel and R.~Gravina, IBVS, No.\,2750 (1985).

\item  S.D.~Friedman, D.G.~York, B.J.~McCall, et al., Astrophys. J. \textbf{727} 33 (2011).

\item  A.S.~Miroshnichenko, private communication.

\item  Th.~Neckel, G.~Klare, \& M.~Sarcander, Astron. \& Astrophys. Suppl. \textbf{42} 251 (1980).

\item  J.~Brand \& L.~Blitz, Astron. \& Astrophys. \textbf{275} 67 (1993).

\item  R.M.~Humphreys, Astrophys. J. Suppl. \textbf{38} 309 (1978).

\item G.~M\"unch, Astrophys. J. \textbf{125} 42 (1957).

\end{enumerate}

\newpage

\begin{longtable}{l|l @{{\vrule width 1pt}} c | c @{{\vrule width 1pt}} c|c  @{{\vrule width 1pt}} c @{{\vrule width 1pt}} r|r @{{\vrule width 1pt}} r | r}
\caption{Central residual intensities and heliocentric radial velocities for the lines in the MWC\,17 spectrum. Designations
         are the same as in Table\,2. Uncertain measurements are marked with a colon}
\label{Ident}         
\\ \hline          
Ident            & $\lambda$ & r    &\hspace{0.2cm} Vr\hspace{0.2cm} &\multicolumn{2}{c @{{\vrule width 1pt}}}{V$_{em}$} & \hspace{0.2cm}V$_{abs}$ \hspace{0.2cm}
                 &\multicolumn{2}{c  @{{\vrule width 1pt}}}{V(r$/$2} & \multicolumn{2}{c}{V(r$\approx$1)} \\
\endfirsthead
\hline
Ident     & $\lambda$, \AA{} & \hspace{0.2cm} r \hspace{0.2cm}  &\hspace{0.2cm} Vr \hspace{0.2cm} &\multicolumn{2}{c  @{{\vrule width 1pt}}}{V$_{em}$} 
          &\hspace{0.2cm} V$_{abs}$ \hspace{0.2cm} &\multicolumn{2}{c @{{\vrule width 1pt}} }{V(r$/$2)} & \multicolumn{2}{c}{V(r$\approx$1)} \\
\hline    
\endhead
\hline
\multicolumn{11}{r}{Table\,5, to be continued } \\ 
\hline
\endfoot
\hline
\endlastfoot 
\hline                 
${\rm[SII]}$1F   & 4068.60   & 4.6  &  $-48$:  &           &         &         &  $-77$: & $-22$:   & $ -98$: &  $-5$:  \\
${\rm[SII]}$1F   & 4076.35   & 2.7: &  $-47$:  &           &         &         &         &          &         &        \\
H$\delta$        & 4101.74   & 5.9  &          &           &  $-23$  &  $-61$: &         &          &         &        \\
FeII(28)         & 4178.85   & 2.7  &  $-48 $  &    $-67$: &  $-24$: &  $-46$  &  $-90$: & $-7$:    & $-100$: &     7: \\
FeII(27)         & 4233.17   & 4.8  &  $-50 $  &    $-68$: &  $-35$  &  $-49$  &  $-84$: & $-21$:   & $-105$: &    10  \\
${\rm [FeII]}$21F& 4243.98   & 2.6: &  $-47$:  &           &         &         &         &          &         &        \\
${\rm [FeII]}$21F& 4276.83   & 3.1: &  $-50 $  &           &         &         &         &          &         &        \\
${\rm [FeII]}$7F & 4287.39   & 7.1  &  $-48 $  &    $-57$  &  $-42$  &  $-52$  &  $-72$: & $-26$:   & $-95$:  &      5:\\
${\rm [FeII]}$21F& 4319.62   & 2.1  &  $-48$:  &           &         &         &         &          &         &        \\
H$\gamma$        & 4340.47   &14.0  &          &           &  $-21$: &  $-58$  &         &          &         &        \\
${\rm [FeII]}$7F &  4359.33  &5.6   &  $-46 $  &           &         &         &         &  $-25$:  &         &     0:  \\
FeII(32)         &  4413.59  &      &          &           &         &         &         &          &         &         \\
${\rm [FeII]}$7F &  4413.78  &4.6   &  $-50 $  &           &         &         &  $-74$: &  $-27$:  & $-111$: &  $-3$:  \\
${\rm [FeII]}$6F &  4416.27  &4.3   &          &           &         &         &         &          &         &         \\
${\rm [FeII]}$7F &  4452.10  &3.4   &  $-48 $  &   $ -57$: &  $-40$  & $-50$:  & $ -74$: &  $-26$:  & $ -98$: &  $-4$: \\
${\rm [FeII]}$6F &  4457.94  &2.6   &  $-47 $  &   $ -60$: &  $-40$: & $-52$:  & $ -74$: &  $-28$:  & $-102$: &  $-7$: \\
HeI(14)          &  4471.52  &2.3   &  $-48$:  &   $ -82$: &  $-29$: & $-59$:  & $-105$: &          & $-142$: &        \\
${\rm [FeII]}$7F &  4474.90  &2.4   &  $-49$:  &           &         &         & $ -80$: &  $-22$:  & $-108$: &  $-3$:  \\
FeII(37)         &  4491.40  &2.0:  &  $-53$:  &   $ -73$: &   $-31$:&  $-52$: & $ -92$: &          & $-105$: &        \\
FeII(38)         &  4508.28  &2.0   &  $-50 $  &   $ -72$: &   $-37$ &  $-51$: &         &          & $ -92$: &    8: \\
FeII(37)         &  4515.33  &3.4   &  $-51 $  &   $ -67 $ &   $-28$:&  $-49$: &  $-87$: &  $-18$:  &         &         \\
FeII(37)         &  4520.22  &3.4   &  $-50 $  &   $ -69$: &   $-27$:&  $-45$: &  $-89$: &  $ -8$:  & $-106$: &         \\
FeII(38)         &  4522.63  &3.4   &  $-51 $  &   $ -65 $ &   $-33$:&  $-49$: &  $-81$: &  $-19$:  & $-102$: &     7   \\
FeII(38)         &  4549.47  &3.1   &  $-51 $  &   $ -70$: &   $-32$:&  $-50$: &  $-87$: &  $-15$:  & $-107$: &     0:  \\
TiII(82)         &  4549.63  &      &          &           &         &         &         &          &         &         \\
FeII(37)         &  4555.89  &3.6   &  $-50 $  &   $-72$:  &   $-30$ &  $-45$  &  $-89$: &  $-13$:  & $-107$: &     2:  \\
FeII(38)         &  4576.33  &2.1   &  $-50$:  &           &         &         &         &          &         &         \\
FeII(38)         &  4583.83  &6.5   &  $-49 $  &   $ -70 $ &   $-35$ &  $-51$: &  $-87$  &  $-10$   &         &    13:  \\
FeII(37)         &  4629.33  &6.0   &  $-50 $  &   $ -71 $ &   $-35$:&  $-51$: &  $-88$  &  $-10$:  & $-109$: &    12:  \\
${\rm [FeII]}$4F &  4639.67  &2.5   &  $-49$:  &           &   $-47$:&         &  $-75$: &  $-28$:  & $ -90$: &   $-10$:  \\
${\rm [FeIII]}$3F&  4658.1   & 7.0  &  $-46 $  &   $ -67$: &   $-21$ &  $-49$ :&  $-91$: &  $ -7$:  & $-122$: &    16   \\
FeII(37)         &  4666.75  & 2.0  &  $-49$:  &   $ -68$: &   $-22$:&         &         &          & $-123$: &     3:  \\
${\rm [FeIII]}$3F&  4701.5   & 2.6: &  $-46$:  &   $ -68$: &   $-17$:&  $-45$: &  $-92$: &  $ -4$:  & $-108$: &    14:  \\
HeI(12)          &  4713.18  &1.5:  &  $-49$:  &   $-88: $ &   $-20$:&  $-52$: &         &          &         &         \\
${\rm [FeII]}$4F &  4728.07  &4.0:  &  $-46 $  &   $-53: $ &   $-42$:&  $-48$: &  $-69$: &  $-26$:  & $ -94$: &     0:  \\
FeII(43)         &  4731.47  &1.7:  &  $-51$:  &           &         &         &         &          &  $-95$: &  $-4$:  \\
${\rm [FeIII]}$3F&  4733.9   &1.8:  &  $-48$:  &   $-60: $ &   $-24$:&  $-42$: &         &          &         &         \\
${\rm [FeIII]}$3F&  4754.7   &2.0   &  $-48$:  &   $-75: $ &   $-23$:&  $-55$: &         &          & $-112$: &    20:  \\
${\rm [FeIII]}$3F&  4769.4   &1.7:  &  $-43$:  &   $-63: $ &   $-18$:&  $-44$: &         &          & $ -95$: &    10:  \\
${\rm [FeII]}$20F&  4774.72  &2.2   &  $-48 $  &           &         &         &  $-69$: &   $-25$: & $ -94$: &  $-8$:  \\
${\rm [FeIII]}$3F&  4777.7   &1.3:  &  $-51$:  &   $-65: $ &   $-23$:&   $-42$:&         &          & $-114$: &    23:  \\
${\rm [FeII]}$20F&  4814.53  &4.8:  &  $-46 $  &   $-56: $ &   $-42$:&   $-52$:&  $-75$: &   $-24$: & $-102$: &    5:   \\
CrII(30)         &  4824.14  &1.3:  &  $-46$:  &   $-69: $ &   $-23$:&   $-50$:&         &          & $-104$: &    9:   \\
H$\beta$         &  4861.33  &12.0: &          &           &         &   $-96$:&         &          &         &         \\
                 &           &      &          &           &   $-15$ &   $-56$ &         &          &         &         \\
${\rm [FeII]}$20F&   4874.48 &2.1   &  $-47 $  &           &         &         &  $-74$: &  $-24$:  & $ -90$: &    0:   \\
${\rm [FeII]}$4F &   4889.62 &4.8   &  $-47 $  &   $-54$:  & $-42$:  &  $-50$: &  $-70$: &  $-26$:  & $ -98$: &   $-1$:   \\
${\rm [FeII]}$   &   4898.61 &2.2   &  $-48 $  &           &         &         &         &          & $ -90$: &  $-16$:    \\
${\rm [FeII]}$20F&   4905 34 &3.0   &  $-47 $  &   $-53$:  & $-41$:  &  $-49$: &  $-73$: &  $-23$:  & $ -98$: &    0:   \\
HeI(48)          &   4921.93 &2.1:  &  $-51$:  &  $-101$:  & $-25$:  &         &         &          & $-115$: &        \\
FeII(42)         &   4923.92 &6.0   &  $-43$:  &  $-55  $  & $-36$   &  $-50$  &  $-75$  &  $-17$   & $-103$: &    8:   \\
${\rm [FeIII]}$1F&   4930.5  &1.7   &  $-42$:  &   $-69$:  & $-18$:  &  $-55$: &  $-83$: &  $  5$:  & $-106$  &   25:   \\
${\rm [FeII]}$20F&   4947.37 &1.8   &  $-45$:  &           &         &         &         &          & $ -80$: &         \\
${\rm [FeII]}$20F&   4950.74 &1.9   &  $-45 $  &           &         &         & $ -64$: &  $-27$   & $ -83$: &   $-8$:  \\
${\rm [OIII]}$1F &   4958.92 &2.0   &  $-48$:  &           &         &         & $-127$: &  $ 21$:  & $-170$: &    90:  \\
${\rm [FeII]}$20F&   4973.39 &2.1   &  $-48 $  &  $ -58$:  & $-40$:  &  $-47$: & $ -70$: &  $-26$   & $ -90$: &   $-8$:  \\
${\rm [OIII]}$1F &   5006.84 &4.0   &  $-56$:  &  $-100$:  & $-50$:  &  $-78$: & $-130$: &  $ 19$:  & $-183$: &    95:  \\
${\rm [FeIII]}$1F&   5011.3  &2.5   &  $-46 $  &  $ -60$:  & $-22$   &  $-48$: & $ -92$: &  $ -9$:  & $-119$: &    20:  \\
HeI(4)           &   5015.68 &3.6   &  $-47$:  &  $ -73$:  & $-23$   &  $-46$: &         &  $ -2$:  & $-120$: &    20:  \\
FeII(42)         &   5018.44 &6.5   &  $-43 $  &  $ -55$   & $-37$   &  $-51$: &  $-72$: &  $-20$   & $-108$: &     6: \\
${\rm [FeII]}$20F&   5020.23 &2.0:  &  $-45$:  &           & $-40$:  &  $-46$: &         &          &         &         \\
FeII             &   5030.64 &1.8:  &  $-45$:  &  $ -62$:  & $-37$   &  $-53$: &  $-70$: &  $-22$:  & $  -93$:&   $-5$:  \\
SiII(5)          &   5041.03 &2.1   &  $-46 $  &           &         &         &  $-78$: &  $-18$:  & $ -110$:&      2: \\
${\rm [FeII]}$20F&   5043.52 &1.6   &  $-46 $  &           &         &         &         &          & $  -85$:&  $-11$:  \\   
HeI(47)          &   5047.74 &      &          &           &         &         &         &          &         &         \\
${\rm [TiII]}$19F&   5047.91 &2.0   &  $-45$:  &           &         &         &         &          &         &         \\
SiII(5)          &   5056.06 &2.8   &  $-46 $  &           & $-36$:  &  $-49$: &  $-95$: &   $-5$:  &  $-130$:&    20: \\
${\rm [FeII]}$   &   5060.08 &1.5   &  $-47 $  &           &         &         &  $-72$: &   $-27$: &  $-82$: &  $-18$:\\
FeII             &   5070.90 &1.5   &  $-48$:  &           &         &         &         &          &         &         \\
${\rm [FeII]}$19F&   5072.39 &1.3:  &  $-47$:  &           &         &         &         &          &         &         \\
FeII             &   5075.77 &1.3:  &  $-43$:  &           &         &         &         &          &         &         \\
FeII             &   5089.22 &1.5   &  $-48$:  &   $-68$:  & $-42$:  & $-51$:  &  $-83$: &   $-17$: &  $ -98$:&      0:  \\
FeII             &   5100.74 &1.3   &  $-46$:  &           &         &         &  $-86$: &   $-23$: &  $-105$:&   $-4$:  \\
${\rm [FeII]}$18F&   5107.94 &1.7   &  $-49 $  &           &         &         &  $-80$: &   $-28$: &  $ -96$:&   $-12$:   \\
${\rm [FeII]}$19F&   5111.63 &2.0   &  $-47 $  &   $-61$:  & $-42$   &         &  $-75$: &   $-22$: &  $ -92$:&   $-7$:  \\
FeIII(5)         &   5114.10 &1.7   &  $-44 $  &           &         &         &         &          &         &          \\
FeIII            &   5149.33 &      &          &           &         &         &         &          &         &          \\
FeII             &   5149.46 &1.5   &          &           &         &         &         &          &         &          \\
${\rm [FeII]}$18F&   5158.00 &      &          &           &         &         &         &          &         &          \\
${\rm [FeII]}$19F&   5158.78 &5.5   &  $-50 $  &  $ -55$:  & $-42$:  & $-48$:  & $-102$: &   $-21$  &         &     4: \\
${\rm [FeII]}$35F&   5163.95 &2.4   &  $-46 $  &  $ -53$:  & $-40$:  & $-49$:  & $ -68$  &   $-26$  &  $ -85$:&    $-4$:  \\
FeII(42)         &   5169.03 &6.0   &  $-42 $  &  $ -55 $  & $-36$   & $-50$:  & $ -70$: &   $-17$  &  $-100$:&      7:  \\
${\rm [FeII]}$18F&   5181.95 &1.9   &  $-46 $  &           &         &         & $ -72$: &   $-28$: &  $ -90$:&   $-10$:   \\
FeII(49)         &   5197.58 &4.0   &  $-50 $  &  $ -72 $  & $-33$   & $-52$:  & $ -89$  &   $-10$: &  $-109$:&       3: \\
${\rm [FeII]}$35F&   5199.17 &1.9   &  $-48$:  &           &         &         &         &   $-23$: &         &   $-4$: \\
FeII             &   5203.64 &1.4   &  $-47 $  &           & $-45$:  & $-58$:  & $ -78$: &   $-16$: &   $-90$:&   $-4$: \\
FeII             &   5216.85 &1.8   &  $-45$:  &           &         &         & $ -74$: &   $-19$: &         &   $-5$:\\
${\rm [FeII]}$19F&   5220.06 &2.1   &  $-46 $  &  $ -53$:  & $-41$:  & $-48$:  & $ -71$: &   $-24$: & $  -93$:&   $-6$: \\
FeII(49)         &   5234.62 &4.2   &  $-50 $  &  $ -71$   & $-33$   & $-52$:  & $ -90$  &   $- 9$  & $ -106$:&     10:  \\
FeII             &   5247.95 &1.4   &  $-47$:  &           &         &         & $ -84$: &   $-13$: & $  -97$:&     13:  \\
FeII             &   5251.23 &1.2   &  $-47$:  &           &         &         &         &          & $  -96$:&       0: \\
FeII(49)         &   5254.93 &1.5:  &  $-54$:  &  $ -72$:  & $-31$:  & $-50$:  &         &          &  $-114$:&   $-4$:  \\
${\rm [FeII]}$19F&   5261.62 &4.5   &  $-46 $  &  $ -54$   & $-42$   & $-50$:  & $ -74$: &  $-22$:  &  $ -99$:&     8: \\
FeII             &   5264.18 &      &          &           &         &         &         &          &         &        \\
MgII917)         &   5264.22 &2.0   &  $-45$:  &           &         &         &         &          &         &          \\
FeII(48)         &   5264.80 &      &          &           &         &         &         &          &         &          \\
${\rm [FeII]}$18F&   5268.87 &2.4   &  $-46$:  &           &         &         &         &          &         &          \\
${\rm [FeIII]}$1F&   5270.40 &4.3   &  $-42 $  &  $ -62$:  & $-15$   & $-45$:  &         & $ -4$    &         &   18:  \\
${\rm [FeII]}$18F&   5273.35 &4.9   &  $-47 $  &  $ -52$:  & $-41$:  & $-50$:  &  $-70$  & $-27$    & $  -98$:&      0:  \\
FeII(49)         &   5276.00 &6.5   &  $-50 $  &  $ -70 $  & $-32$:  & $-47$:  &  $-88$  & $-10$    & $ -112$:&      9:  \\
FeII(41)         &   5284.10 &4.0   &          &           &         &         &         &          &         &          \\
FeII             &   5291.67 &1.9   &  $-47 $  &  $ -60$:  & $-33$:  & $-51$:  &  $-78$: & $-15$:   & $ -100$:&     2:   \\
${\rm [FeII]}$19F&   5296.83 &2.0   &  $-47 $  &           &         &         &  $-74$  & $-26$:   & $  -92$:&   $-10$:  \\
FeII(49,48)      &   5316.66 &2.0   &  $-51 $  &  $-69 $   & $-35$   & $-51$   &  $-88$  & $-13$    & $ -114$:&      13: \\
FeII(49)         &   5325.56 &1.8   &  $-50 $  &  $-75$:   & $-33$:  &         &  $-90$: & $ -3$:   & $ -103$:&      3:  \\
OI(12)           &   5330.74 &1.4   &  $-49$:  &           &         &         &         &          &         &          \\
${\rm [FeII]}$19F&   5333.65 &4.0   &  $-47 $  &           & $-42$:  &         &  $-72$  & $-25$    &  $ -95$ &       0: \\
FeII(48)         &   5362.87 &3.0   &  $-51 $  &   $-73$   & $-35$:  & $-48$:  &  $-91$  & $-10$:   &  $-108$:&       8: \\
${\rm [FeII]}$19F&   5376.45 &3.6   &  $-47 $  &   $-54$:  & $-42$   & $-51$:  &  $-71$  & $-27$    &  $-100$:&   $-2$: \\
FeII             &   5402.06 &1.5   &  $-43$:  &           &         &         &  $-73$: & $-20$    &  $-100$:&       5: \\
${\rm [FeII]}$17F&   5412.65 &2.3   &  $-49 $  &   $-54$:  & $-44$:  & $-49$:  &  $-77$: & $-26$    &  $-105$:&   $-2$: \\
FeII(48)         &   5414.05 &      &          &           &         &         &         &          &         &          \\
FeII(49)         &   5425.25 & 2.1  &  $-51 $  &   $-73$:  &         &         &  $-91$  & $-10$    &  $-106$:&     10   \\
FeII             &   5427.82 & 1.2  &   $-52$: &           &         &         &         &          &         &          \\
FeII(55)         &   5432.98 & 2.5: &          &           &         &         &         &          &         &          \\
${\rm [FeII]}$18F&   5433.13 & 1.7: &   $-48$: &           &         &         &         &          &         &          \\
FeII             &   5457.72 & 1.4  &   $-47 $ &           &         &         &  $-75$: &  $-21$:  &  $ -89$:&   $-3$:  \\
${\rm [FeII]}$   &   5477.24 & 1.8  &   $-47 $ &           &         &         &  $-71$: &  $-25$:  &  $ -83$:&   $-17$: \\
FeII             &   5482.31 & 1.5  &   $-51$: &   $-78$:  & $-34$:  &         &  $-87$: &  $-18$:  &  $ -98$:&   $-5$:  \\
FeII             &   5487.63 & 1.3  &   $-50$: &           &         &         &         &          &         &          \\
${\rm [FeII]}$17F&   5495.82 & 1.5  &   $-46 $ &           &         &         &  $-69$: &  $-21$:  &  $ -80$:&   $-4$:  \\
${\rm [FeII]}$17F&   5527.34 & 3.3  &   $-47 $ &   $-53$:  & $-42$:  & $-50$:  &  $-72$  &  $-24$   &  $ -92$:&   $-6$:  \\
FeII(55)         &   5534.83 & 4.0  &   $-50 $ &   $-75$   & $-35$:  & $-45$:  &  $-90$  &  $ -9$   &  $-106$:&     10: \\
${\rm [FeII]3}$9F&   5551.31 & 1.4  &   $-45 $ &           &         &         &  $-69$: &          &  $ -85$:&          \\
OI(24)           &   5554.95 & 1.4  &   $-46$: &           &         &         &         &          &         &          \\
${\rm [OI]}$3F   &   5577.34 & 2.3  &   $-45 $ &   $-69$:  & $-33$:  & $-51$:  &  $-85$: &  $ -4$:  &  $ -98$:&      10: \\    
${\rm [FeII]}$39F&   5588.15 &   1.4&   $-47$: &           &         &         &         &          &  $ -98$:&   $-3$:  \\
${\rm [FeII]}$   &   5613.27 &   1.3&   $-47$: &           &         &         &         &          &         &           \\
${\rm [FeII]}$   &   5673.21 &   1.6&   $-47 $ &   $-55$:  & $-40$:  & $-49$:  &  $-75$: &  $-26$   &         &   $-10$:  \\
${\rm [FeII]}$17F&   5745.70 &      &          &           &         &         &         &          &         &           \\
${\rm [FeII]}$34F&   5746.97 &   2.6&   $-47$  &   $-54$:  & $-41$:  &         &         &  $-28$   &         &   $-5$:  \\
${\rm [NII]}$3F  &   5754.64 &  17.0&   $-47$  &   $-65$   & $-32$   & $-53$   &  $-84$  &  $-15$   &  $-122$:&     8:  \\
FeII             &   5780.12 &  1.4:&   $-48$: &           &         &         &         &          &         &           \\
DIB              &   5780.37 &   0.7&   $  5$: &           &         &         &         &          &         &           \\
FeII             &   5783.63 &   1.4&   $-46$: &           &         &         &         &          &         &           \\
DIB              &   5796.96 &   0.8&   $ -8$: &           &         &         &         &          &         &           \\
FeII(58)         &   5835.49 &   1.7&   $-51$: &           &         &         &         &          &         &           \\
HeI(11)          &   5875.72 &   7.5&   $-52$  &   $-75$:  & $-28$   & $-54$:  & $-106$: &  $-7$:   &  $-145$:&   28:   \\
FeII             &   5885.02 &  1.3:&  $-41$: &           &         &         &         &          &         &           \\
NaI(1)           &   5889.95 &   0.6&   $-59$: &           &         &         &         &          &         &           \\
                 &           &   0.2&   $-50$  &           &         &         &         &          &         &           \\
                 &           &   0.3&   $-38$  &           &         &         &         &          &         &           \\
                 &           &   0.1&   $-27$  &           &         &         &         &          &         &           \\
                 &           &   0.1&   $-10$  &           &         &         &         &          &         &           \\
NaI(1)           &   5895.92 &   0.7&   $-60$: &           &         &         &         &          &         &           \\
                 &           &   0.3&   $-50$  &           &         &         &         &          &         &           \\
                 &           &   0.4&   $-40$  &           &         &         &         &          &         &           \\
                 &           &   0.1&   $-27$  &           &         &         &         &          &         &           \\
                 &           &   0.1&   $ -9$  &           &         &         &         &          &         &           \\
${\rm [FeII]}$34F&   5901.26 &  1.2:&  $-42$: &           &         &         &         &          &         &           \\
FeII             &   5902.83 &   1.4&   $-48$: &   $-74$:  & $-36$:  &         &         &          &  $-89$: &  $-9$:     \\
SiII(4)          &   5957.56 &   3.3&          &           &         &         &         &          &         &           \\
OI(23)           &   5958.6  &      &          &           &         &         &         &          &         &           \\
FeII             &   5965.63 &   1.2&   $-47$: &           &         &         &         &          &         &           \\
SiII(4)          &   5978.93 &   4.3&   $-47$: &   $-57$:  & $-37$:  & $-51$:  &  $-86$: &  $-15$   & $-118$: &   10:     \\
OI(22)           &   6046.4  &   4.2&          &           &         &         &         &          &         &           \\
FeII(46)         &   6113.32 &   1.2&   $-47$: &           &         &         &         &          &         &           \\
CaI(3)           &   6122.22 &      &          &           &         &         &         &          &         &           \\
MnII(13)         &   6122.44 &   1.3&   $-46$: &           &         &         &         &          &         &           \\
MnII(13)         &   6122.80 &      &          &           &         &         &         &          &         &           \\
MnII(13)         &   6123.16 &      &          &           &         &         &         &          &         &           \\
${\rm [TiII]}$22F&   6124.57 &      &          &           &         &         &         &          &         &           \\
SiI(30)          &   6125.02 &   1.9&          &           &         &         &         &          &         &           \\
FeII(74)         &   6147.74 &   1.4&          &           &         &         &         &          &         &           \\       
FeII(74)         &   6149.25 &      &          &           &         &         &         &          &         &           \\             
OI(10)           &   6155.98 &      &          &           &         &         &         &          &         &           \\
OI(10)           &   6156.77 &      &          &           &         &         &         &          &         &           \\
OI(10)           &   6158.18 &   1.4&:  $-50$: &           &         &         &         &          &         &           \\
${\rm [FeII]}$44F&   6188.55 &   1.3&:  $-47$: &           &         &         &         &          &         &           \\
DIB              &   6195.96 &   0.8&   $-12$: &           &         &         &         &          &         &           \\
FeII             &   6233.53 &   2.1&   $-50 $ &           &         &         &  $-88$: &  $-14$:  & $-100$: &   9:      \\
FeII(74)         &   6238.39 &  1.5:&   $-50$: &           &         &         &         &          &         &           \\
FeII(74)         &   6239.90 &      &          &           &         &         &         &          &         &           \\
FeII(74)         &   6247.55 &  2.0:&          &           &         &         &         &          &         &           \\
FeII             &   6248.89 &   3.7&          &           &         &         &         &          &         &  $-13$    \\
DIB              &   6283.85 &   0.8&          &           &         &         &         &          &         &           \\
FeII             &   6291.83 &   1.7&   $-49$: &           &         &         &  $-85$: &  $-12$:  & $-100$: &   6:      \\
${\rm [OI]}$1F   &   6300.30 &  30.0&   $-48$  & $-58$     &  $-39$  & $-50$   &  $-74$  &  $-23$   & $ -93$  &  $-2$     \\ 
${\rm [SIII]}$3F &   6312.10 &  4.6 &   $-48$: & $-62$:    &  $-49$: & $-26$:  &   $-80$:&  $-12$:  & $-118$: &    5:     \\
FeII             &   6317.99 &  8.8 &   $-50$  & $-61$:    &  $-36$  & $-50$:  &   $-87$ &  $-14$   & $-105$: &    8:     \\
SiII(2)          &   6347.10 &  4.5 &   $-47$  & $-56$:    &  $-38$: & $-51$:  &   $-88$ &  $-15$   & $-120$: &  15:      \\
FeII             &   6357.17 &  1.3 &   $-52$: &           &         &         &         &          &         &           \\
${\rm [OI]}$1F   &   6363.78 & 12.0 &   $-48$  & $-58$     &   $-39$ & $-49$   & $-74$   &  $-24$   &  $-93$: &  $-5$:    \\
FeII(40)         &   6369.46 &      &          &           &         &         &         &          &         &           \\
SiII(2)          &   6371.36 &  3.0 &   $-47$  &           &         &         & $-84$:  &  $-18$:  &  $-120$:&     5:    \\
FeII             &   6375.79 &  1.2 &   $-49$  &           &         &         &         &          &  $ -90$:&  $-5$:    \\
FeII             &   6383.72 &  5.1 &   $-51$  & $-62$:    &   $-37$:& $-49$   & $-86$:  &          &  $-105$:&          \\
FeII             &   6385.45 &  3.3 &   $-48$: & $-64$:    &   $-34$:& $-47$:  &         &  $-16$   &         &      7:   \\
FeII(74)         &   6416.92 &  1.6 &   $-52$: & $-74$     &   $-28$:& $-42$:  & $-87$:  &          &  $-102$:&      0:   \\
FeII(40)         &   6432.68 &  1.5 &   $-52$: &           &         &         &         &          &  $-115$:&    10:    \\ 
FeII             &   6442.95 &  2.6 &   $-50$  & $-60$:    &   $-35:$&  $48$   & $-86$:  &  $-12$:  &  $-100$:&     3:    \\
FeII             &   6455.84 &  3.5 &          &           &         &         &         &          &         &           \\
FeII(74)         &   6456.38 &  2.6 &          &           &         &         &         &          &         &           \\
FeII             &   6491.25 &  2.1 &   $-49$: &           &         &         & $-84$:  &          &  $-99$: &          \\
FeII             &   6493.03 &  2.6 &   $-51$  &           &         &         &         &  $-14$:  &         &     5:    \\
${\rm [NII]}$1F  &   6548.03 &  1.5 &   $-47$  &           &         &         & $-85$:  &  $-13$:  &  $-115$:&   10:     \\
H$\alpha$        &   6562.81 &      &   $-52$: &           & $-12$   & $-108$: &         &          &         &           \\
                 &           &      &          &           &         & $-54$   &         &          &         &           \\
${\rm [NII]}$1F  &   6583.45 &  3.4 &   $-46$  &           &         &         & $-82$   &  $-10$   &  $-115$:&    15:    \\
FeII             &   6586.70 &  1.3 &   $-50$: &           &         &         & $-86$:  &  $-12$:  &  $-102$:&    7:     \\
DIB              &   6613.56 &  0.8 &   $-19$: &           &         &         &         &          &         &           \\
HeI(46)          &   6678.15 &  2.5 &   $-53$  & $-72$:    & $-26$   & $-60$:  & $-105$: &  $ -1$:  &  $-135$:&    20:    \\ 
${\rm [SII]}$2F  &   6716.47 &  1.5 &   $-48$  &           &         &         & $-65$:  &  $-27$:  &  $ -80$:&  $-15$:    \\
${\rm [SII]}$2F  &   6730.85 &  2.0 &   $-48$  &           &         &         & $-71$:  &  $-26$:  &  $ -95$:&   $-5$:    \\
\end{longtable}

\end{document}